# Towards an All-Silicon QKD Transmitter Sourced by a Ge-on-Si Light Emitter

Florian Honz, *Student Member, IEEE*, Nemanja Vokić, *Member, IEEE*, Michael Hentschel, Philip Walther, Hannes Hübel, *Member, Optica*, and Bernhard Schrenk, *Member, IEEE*

*Abstract*—We demonstrate a novel transmitter concept for quantum key distribution based on the polarization-encoded BB84 protocol, which is sourced by the incoherent light of a forward-biased Ge-on-Si PIN junction. We investigate two architectures for quantum state preparation, including independent polarization encoding through multiple modulators and a simplified approach leveraging on an interferometric polarization modulator. We experimentally prove that the Ge-on-Si light source can accommodate for quantum key generation by accomplishing raw-key rates of 2.15 kbit/s at a quantum bit error ratio of 7.71% at a symbol rate of 1 GHz. We further investigate the impact of depolarization along fiber-based transmission channels in combination with the broadband nature of the incoherent light source. Our results prove the feasibility of a fully-integrated silicon quantum key distribution transmitter, including its light source, for possible short-reach applications in zero-trust intra-datacenter environments.

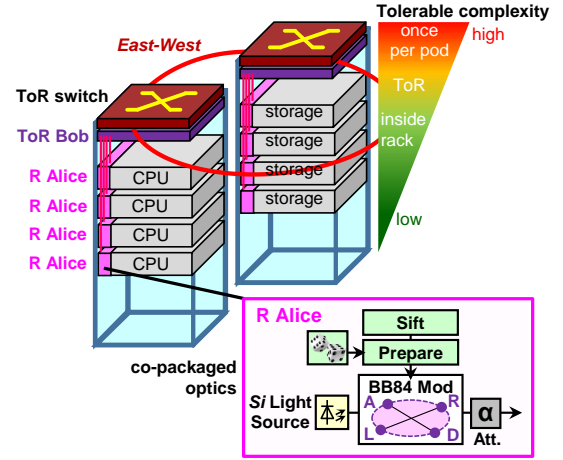

**Fig. 1.** Intra-datacenter application example for intra-rack and intra-pod QKD integration, where distributed low-complexity QKD transmitters (R Alice) are associated to a pooled top-of-rack receiver (ToR Bob). A higher degree of complexity can be tolerated towards the more centralized top of a rack, suitable for a more complex Bob. The inset shows the distributed and simplified QKD transmitters.

## I. INTRODUCTION

The quantum computer is on the verge of entering the application space, with systems having more than 400 Qubits being already developed and tested [1]. The fast advances threaten the security of our current cryptographic systems, which are based on the prime factorization of large numbers, something which is a computationally hard challenge for classical computers. Their quantum-enabled contenders, on the other hand, can solve these problems exponentially faster. This calls for an urgent upgrade of our current cryptographic standards, something that Quantum Key Distribution (QKD) can provide. QKD guarantees absolute security since it does not build on computationally hard problems but on the laws of nature, namely the quantum properties of photons. Commercial QKD devices have been successfully established in the market as a solution that is proven to be secure and compatible with deployments in 19-inch server racks. The latter reflects that nowadays QKD systems are still rather bulky and subject to a considerably large form-factor that hinders the mass deployment, especially in applications where space is precious. With the advent of the zero-trust model, intra-datacenters communications, where server densities reach numbers of 10,000 or more [2], are starting to become a playground for QKD. Such deployment calls for a disruptive miniaturization of QKD devices, while the cost associated to a seamless integration with storage and compute units has to be lowered likewise. Photonic integration efforts concerning QKD subsystems have offered vistas towards these directions [3-28]. Various works have focused on compact form-factors and stable key elements, such as delay interferometers [3, 14, 26]. Other works have demonstrated compact state preparation schemes leveraging silicon integration technology [13-27], however, failing to capture the light source, resulting in complex assembly/packaging for the entire QKD transmitter. A monolithic integration approach on InP has been shown for a QKD transmitter [28], however, though providing inherent optical

This work was supported by the European Research Council (ERC) under the EU Horizon-2020 programme (grant agreement No 804769) and by the Austrian FFG Research Promotion Agency and NextGeneration EU (grant agreement No FO999896209). *(Corresponding author: Florian Honz)*.

F. Honz, N. Vokić, H. Hübel and B. Schrenk are with the AIT Austrian Institute of Technology, Center for Digital Safety & Security, Giefinggasse 4, 1210 Vienna, Austria (e-mail: florian.honz@ait.ac.at).

P. Walther is with the University of Vienna, Faculty of Physics, Vienna Center for Quantum Science and Technology (VCQ), Boltzmanngasse 5, 1090 Vienna, Austria (e-mail: philip.walther@univie.ac.at).

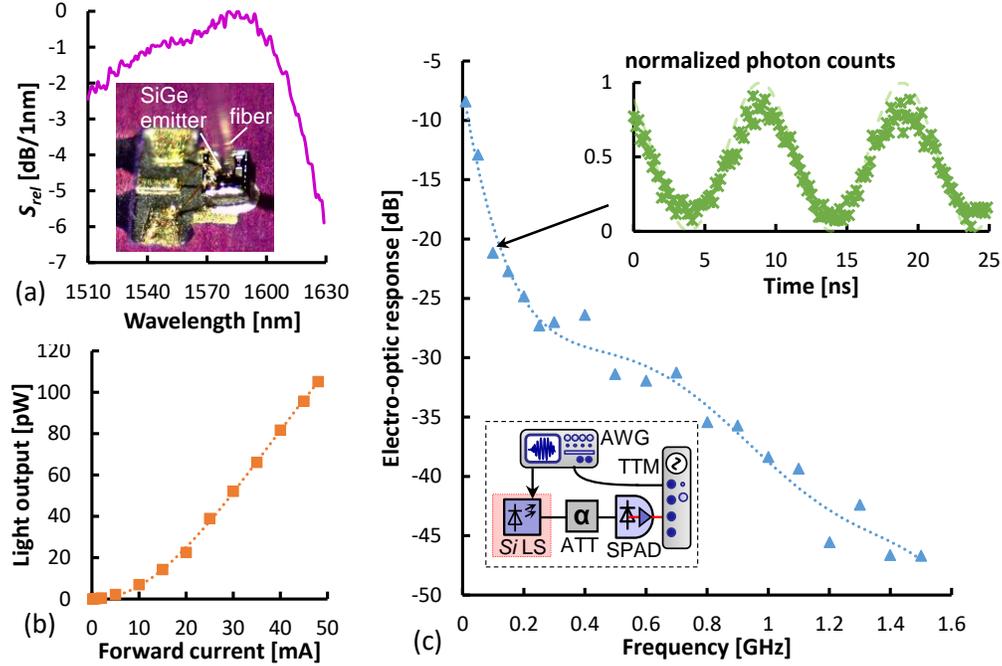

**Fig. 2.** (a) Emission spectrum of the Ge-on-Si light emitter in the C+L bands, (b) L-I characteristics of the Ge-on-Si light emitter, and (c) electro-optic bandwidth of the Ge-on-Si emitter, including the optical output for direct modulation at 100 MHz.

gain, InP is widely incompatible with photonics/electronics co-integration. These earlier works therefore drive to a conclusion that there is no one-fits-all solution when it comes to integration platforms: Especially the inclusion of the light source necessitates hetero-integration schemes that blend optically active III-V materials with other platforms that cater for the needs of high-speed modulation and electronic co-integration, such as silicon.

However, the application of QKD in intra-datacenter environments can build on the unidirectional nature of the quantum channel: There is only one QKD transmitter (Alice) and one QKD receiver (Bob) required, in stark contrast to the bidirectional transceivers of the classical datacenter interconnects. This enables the formation of a QKD link with asymmetric complexity profile for the involved QKD sub-systems, where for example a greatly simplified Alice is distributed in the rack domain while a more complex Bob is centralized at the top-of-the-rack (ToR).

Figure 1 introduces such a concept for the datacenter realm for discrete-variable QKD. Here, Bob's single-photon avalanched detectors (SPADs) are pooled at a ToR location, while multiple Alices, based on simplified QKD transmitters, are distributed within the individual rack units. The key-exchange that is being performed unidirectionally then supplies the key for securing the bidirectional interconnects within the zero-trust datacenter environment. This scheme implies that the main complexity reduction should happen at Alice' QKD transmitter.

In this paper, we extend our recent findings [29] on a suitable methodology aiming at the seamless integration of the QKD transmitter and the switch ASIC according to the paradigm of co-packaged optics for datacenter applications [30]. Towards this direction, we investigated the adoption of a Ge-on-Si light emitter, which can be monolithically integrated on silicon platforms, as the source for a BB84-based polarization-encoded QKD scheme. We experimentally demonstrate that such a system can operate at a QBER of 7.71% and a raw-key rate of 2.15 kb/s, when paired with independently modulated polarization encoders. We further investigate how polarization modulation can be further simplified by introducing an integrated dual-polarization I/Q modulator, paving the way towards an all-silicon QKD transmitter.

The paper is organized as follows. Section II characterizes the proposed Ge-on-Si light emitter while Section III will investigate the QKD performance that can result from the inclusion of such a light emitter in a QKD transmitter with independently modulated polarization states. Section IV will introduce a concept for simplified polarization modulation that is compatible with co-integration of the proposed light source. Section V covers depolarization effects as the main impairment identified for QKD transmission with incoherent light sources. Section VI evaluates the performance limits for a QKD system employing the proposed polarization encoder when seeded with an incoherent light source. Finally, Section VII summarizes our work and concludes with an outlook on possible further improvements.

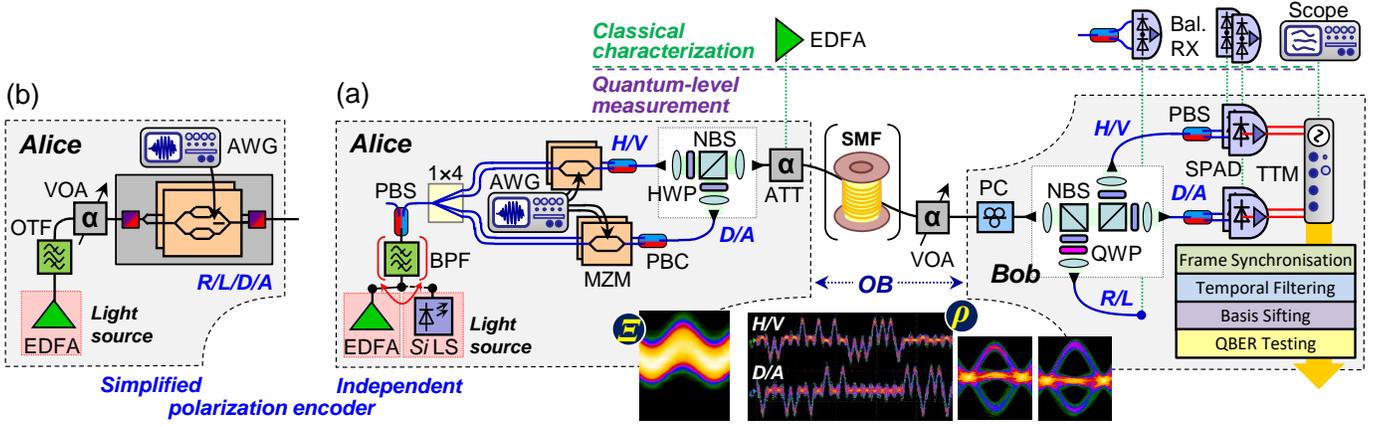

**Fig. 3.** Experimental setup for QKD evaluation with the Ge-on-Si light source, using (a) independent polarization state encoding and (b) simplified polarization encoding at Alice' QKD transmitter. For the latter, the **D/A** and **R/L** bases were used at Bob.

## II. GE-ON-SI LIGHT EMITTER

The employed light emitter can be seen in Fig. 2(a). It is a die-level Ge-on-Si PIN junction which is wire-bonded to a PCB and vertically fiber-coupled. When forward-biasing this PIN junction, it acts as a broadband light source in the C+L bands, with the peak emission showing a range of ~20 nm centered at 1581 nm (Fig. 2(a)). This is essential for telecom applications where the C+L wavebands are relevant due to minimal fiber attenuation. Furthermore, as can be seen in Fig. 2(b), the junction has an LED-like light emission with a maximum output power of 105 pW at a current of 48 mA, corresponding to -69.8 dBm. Due to the forward current limit of the chosen PIN junction, higher currents and thereby an increased light output have not been pursued for this particular sample. Nonetheless, the Ge-on-Si junction is strong enough to source a QKD transmitter since such a transmitter operates at a mean photon number of $\mu_Q = 0.1$ photons/symbol and a symbol rate of 100 MHz therefore relates to a launched output power of -88.9 dBm. This would leave ~19 dB of headroom to accommodate for transmitter losses at Alice. Likewise, a symbol rate of 1 GHz leads to a headroom of ~9 dB. This renders the proposed incoherent light source suitable for QKD, provided that a suitable method for state encoding can be facilitated. This will be investigated shortly in Section III.

Furthermore, we investigated the possibility to directly modulate the PIN junction. Since the emission is very weak and mainly at the single photon level, a conventional measurement of the response with a vector network analyzer was not possible. Therefore, as shown in the inset in Fig. 2(c), we used an arbitrary waveform generator (AWG) to perform a sinusoidal drive of the light emitter and acquired the signal by a SPAD. The resulting photon counts were recorded by a time tagging module (TTM). We then evaluated the contrast in the recorded histogram of the SPAD output for a frequency-swept drive. Figure 2(c) reports the resulting electro-optic response of the Ge-on-Si light emitter, defined as the response in received optical power, represented as SPAD counts, relative to that at continuous-wave emission. It further shows the time-domain response to an exemplary modulation signal with a frequency of 100 MHz. Due to the strong roll-off at low frequencies, which occurs despite 50Ω impedance matching, we used external modulation for the broadband drive of the state preparation; However, pulse carving appears to be feasible through direct forward bias current modulation for the emitter.

## III. QKD PERFORMANCE FOR INDEPENDENT POLARIZATION ENCODING

The Ge-on-Si PIN emitter is a broadband source, which is emitting light in a random polarization. For a polarization based QKD protocol this output needs to be prepared in a well-defined polarization state, in particular horizontal (**H**), vertical (**V**) and antidiagonal (**A**), diagonal (**D**), according to the BB84 protocol for polarization-encoded QKD.

Figure 3(a) shows our experimental setup to evaluate QKD functionality towards this direction. The light source we used for our investigations was either a combination of Erbium-doped fiber amplifier (EDFA), whose amplified spontaneous emission (ASE) spectrum has been shaped towards 1539.1 nm through an optical bandpass filter (BPF) having a 25-GHz bandwidth, or the aforementioned Ge-on-Si PIN light source with an optional 14-nm wide BPF, centered at 1590 nm. To encode the photons in the correct polarization states, the output of either light source was first polarized by a polarization beam splitter (PBS) in order to define a well-defined polarization state and then fed to four Mach-Zehnder Modulators (MZMs) for the purpose of independent polarization encoding. Only one of those MZMs was transmitting photons at any given time and through polarization beam combiners (PBC) and free-space optics, which rotate the **H/V** basis through a half-wave plate (HWP) into the **D/A** basis, the states were encoded in the **H/V** and **D/A** basis – as required to implement the polarization-encoded BB84 QKD protocol. We operated the transmitter at a symbol rate of 1 GHz and integrated pulse carving with the return-to-zero MZM drive to suppress symbol transitions. Before transmitting the photons to Bob, we used an optical attenuator to set the signal launch power to a mean photon number of $\mu_Q = 0.1$ photons/symbol.

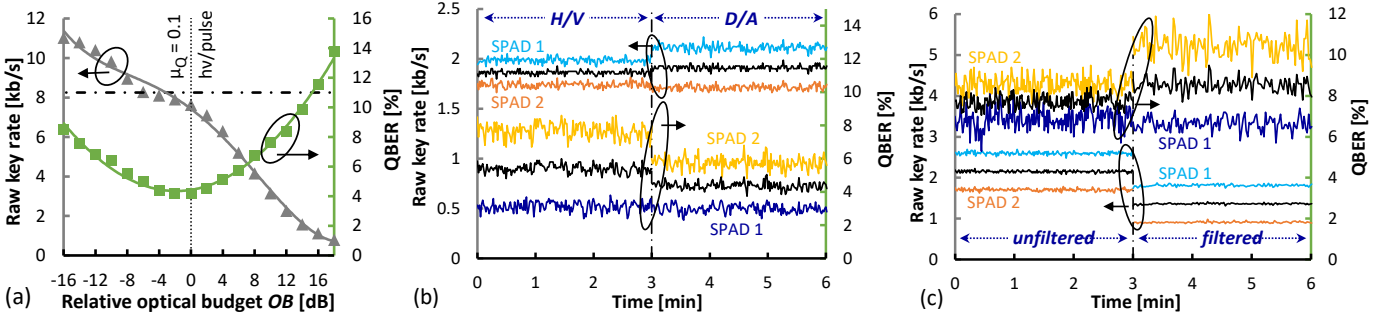

**Fig. 4.** (a) QBER and raw-key rate as a function of the optical budget for the ASE-sourced transmitter and evolution of the raw-key rate and QBER when seeding Alice' polarization encoder with (b) the ASE-sourced and (c) the Ge-on-Si PIN light source.

At Bob's side we manually aligned the polarization of the incoming signal to the axes of our polarization analyzer with a polarization controller (PC). The outputs of the polarization analyzer were connected to a pair of free-running InGaAs SPADs with dark count rates of 560 and 525 cts/s. We consecutively measured the received quantum signal in the **H/V** and **D/A** bases and recorded the detection events of the SPADs with a TTM. A real-time QBER evaluation was then performed, which included frame synchronization with the pre-defined pseudo random bit sequence transmitted by Alice, as well as temporal filtering to 50% of the symbol period to reduce the dark counts. Additionally, basis sifting was performed during this process.

To confirm the correct operation of our transmitter and receiver architectures we also performed a classical characterization. For this we replaced the attenuator by an additional EDFA and the SPADs and TTM with balanced receivers and a real-time oscilloscope. The carved optical output at Alice ($\Xi$) and the characteristic three-level signals ($\rho$) for the **H/V** and **D/A** bases under 4-state polarization encoding are included in Fig. 3(a).

In order to assess the performance as a function of the optical budget (OB) between Alice and Bob, defined as the difference of the transmitted power level at Alice and the received optical power (ROP) at Bob, we used the ASE seed light for Alice' transmitter, together with a variable optical attenuator (VOA) between Alice and Bob to intentionally vary the ROP and investigate link loss effects. The corresponding results are presented in Fig. 4(a). We obtain a raw-key rate (▲) of 7.6 kb/s and a QBER (■) of 4.2% at an OB of 0 dB, which corresponds to $\mu = \mu_Q$. The QBER threshold of 11%, corresponding to the limit at which a secure key can still be generated [31], is reached at an OB of 15.2 dB, meaning that we can erode this power margin through adding (neutral) elements with the equivalent loss between the QKD transmitter and receiver, such as power-splitting signal distributors. From a practical point of view, this means that we could accommodate the loss of an optical splitter connecting up to 16 Alices to one Bob in a datacenter, or even more when resorting to low-loss optical switching technologies [32]. For the sake of completeness, Fig. 4(a) also characterizes the saturation threshold of the SPAD receivers at increased $\mu$-values larger than $\mu_Q$, meaning a negative OB.

The short-term evolution of the QBER and raw-key rates for consecutive measurements in the **H/V** and **A/D** bases over the course of a few minutes can be found in Fig. 4(b). As can be seen in Fig. 3(a) our receiver enabled us to measure the received photons with **H** (**D**) and **V** (**A**) polarization at the same time with different SPADs (SPAD 1 and SPAD 2 in Fig. 4(b)). To calculate the QBERs and raw-key rates from the obtained detection events of the SPADs we used the already described postprocessing stack depicted in Fig. 3(a). On average, we reach a QBER of 5.37% ($3\sigma = 0.78\%$) and 4.28% ($3\sigma = 0.70\%$) per SPAD in the **H/V** and **A/D** basis, respectively, with raw-key rates of 1.87 kb/s and 1.92 kb/s. This proves that our polarization-encoded BB84 transmitter can be operated with incoherent light.

We repeated this measurement using the unfiltered and filtered Ge-on-Si PIN light emitter as seed source for the polarization encoder, reaching raw-key rates of 2.15 kb/s and 1.36 kb/s and QBERs of 7.71% and 8.53%, respectively (Fig. 4(c)). This proves that a light source fully compatible with silicon integration platforms can be applied to potentially realize monolithic integrated QKD transmitters. For this, however, the BB84 polarization encoder should be further simplified.

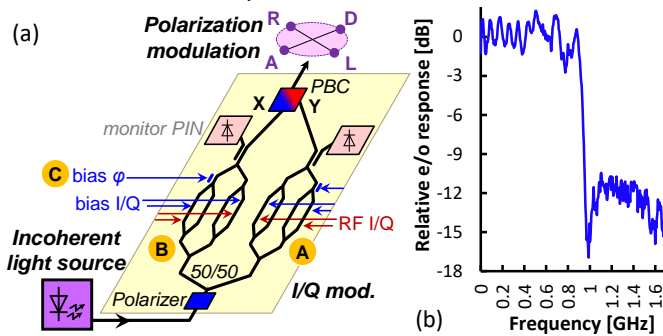

**Fig. 5.** (a) Simplified polarization modulation with a dual-polarization I/Q modulator according to the OIF Implementation Agreement [33] and (b) electro-optic response of its phase section.

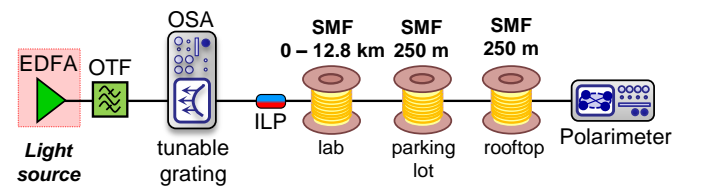

**Fig. 6.** Experimental setup for investigating depolarization.

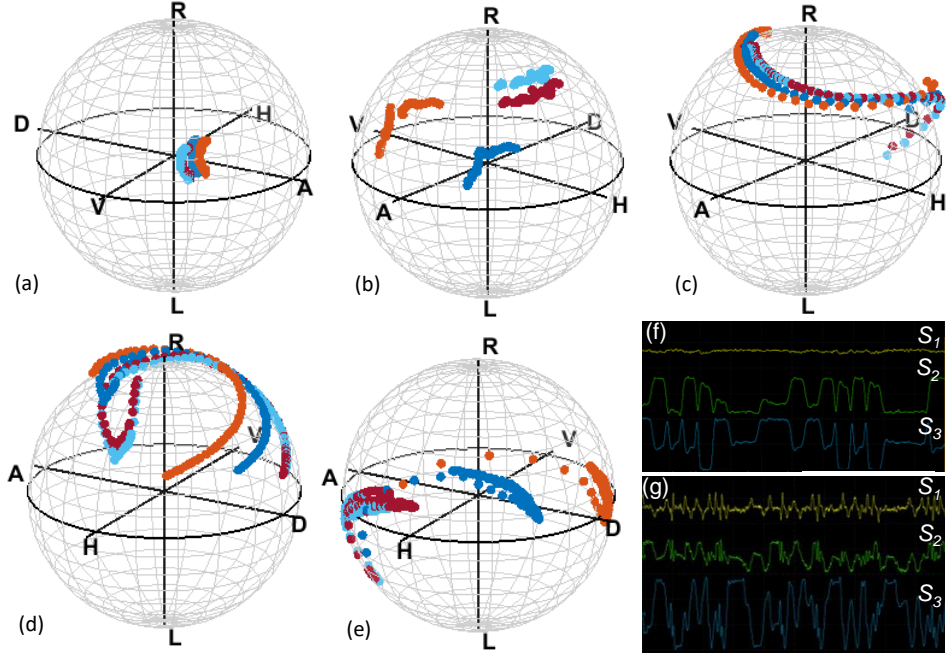

**Fig. 7.** Polarization state evolution for a wavelength of 1570 nm (cyan), 1575 nm (red), 1580 nm (blue) and 1585 nm (orange) in the same fiber over the course of 5 hours for a total fiber length of (a) 500 m, (b) 1.5 km, (c) 4.8 km, (d) 9.1 km and (e) 13.3 km. Fig. (f) and (g) show the effect of depolarization on 1 nm signals being transmitted in the $S_2$ and $S_3$ polarization bases for fiber lengths of (f) 7 m and (g) 12.8 km.

## IV. Simplified Polarization Modulation

As discussed in the previous Section, the preparation of the required polarization states for running a BB84 protocol necessitates an external modulator for the silicon light source. Towards this direction, we propose a more simplified approach through using a dual-polarization in-phase/quadrature (I/Q) modulator, as it is known through the OIF Implementation Agreement [33] for embedding these modulators in metro and long-haul networks. Figure 5(a) introduces this concept. The integrated modulator structure allows the modulation of the **X** and **Y** polarizations through independent data signals to enable optical quadrature amplitude modulation (QAM) in each of the two polarization planes. This QAM is accomplished through nested I/Q modulators (A and B in Fig. 5), whose optical phase $\varphi$ relative to the *I* and *Q* signals can be adjusted separately. Here, we exploit this degree of freedom to perform the desired modulation of the polarization state along the **X/Y** plane of the modulator, which we dedicate to the $S_2/S_3$ plane in the Stokes space: By modulating the phase $\varphi$ (C in Fig. 5) while leaving the *I* and *Q* inputs unmodulated and just use their biases to balance the power in the **X** and **Y** polarizations, we can steer the input polarization state along the $S_2/S_3$ plane and accomplish a 4-state protocol involving the polarizations **A**, **D**, right-circular (**R**) and left-circular (**L**). A prerequisite for this modulation is a suitable electro-optic response for the phase section $\varphi$. Figure 5(b) reports this electro-optic response of the phase section after impedance matching, which has been identified as a prerequisite given the nature of this modulator electrode as a DC bias line. We accomplish a -3 dB bandwidth of 920 MHz, which is sufficiently high in combination with QKD applications having a symbol rate of 1 Gb/s.

## V. Depolarization Effects for Fiber Transmission

The broadband nature of the incoherent seed light necessitates further attention concerning the propagation of the polarization-encoded signal along the transmission fiber. A standard SMF has two distinct and well-defined principal polarization modes with different group velocities [34]. This differential group delay causes polarization mode dispersion (PMD), as known from high-rate transmission systems, but will also exceed the coherence time of the broadband light source. The latter leads to depolarization [35].

This effect has been characterized using a classical polarimeter, together with a sufficiently strong incoherent light source (Fig. 6) based on an L-band EDFA whose ASE has been spectrally defined with a bandwidth ΔΛ through a bandwidth-tunable optical filter (OTF). The output of this filter was then sliced within the spectral bandwidth of the OTF by the tunable grating of an optical spectrum analyzer (OSA), yielding narrow slices with a bandwidth δλ = 1 nm < ΔΛ. These serve as constituent signals for the broadband ensemble ΔΛ. The polarization state of these slices is then defined through an in-line polarizer (ILP) before transmission through three fiber spans: The first had different lengths from 0 to 12.8 km, followed by two 250-m long fiber spans deployed at a parking lot and at a rooftop installation. These latter spans serve a faster evolution of the polarization state through increased exposure to temperature and strain variations. We then measured the state of polarization for each wavelength slice over the course of 5 hours by sweeping the OSA grating filter within ΔΛ during the 5-min measurement intervals.

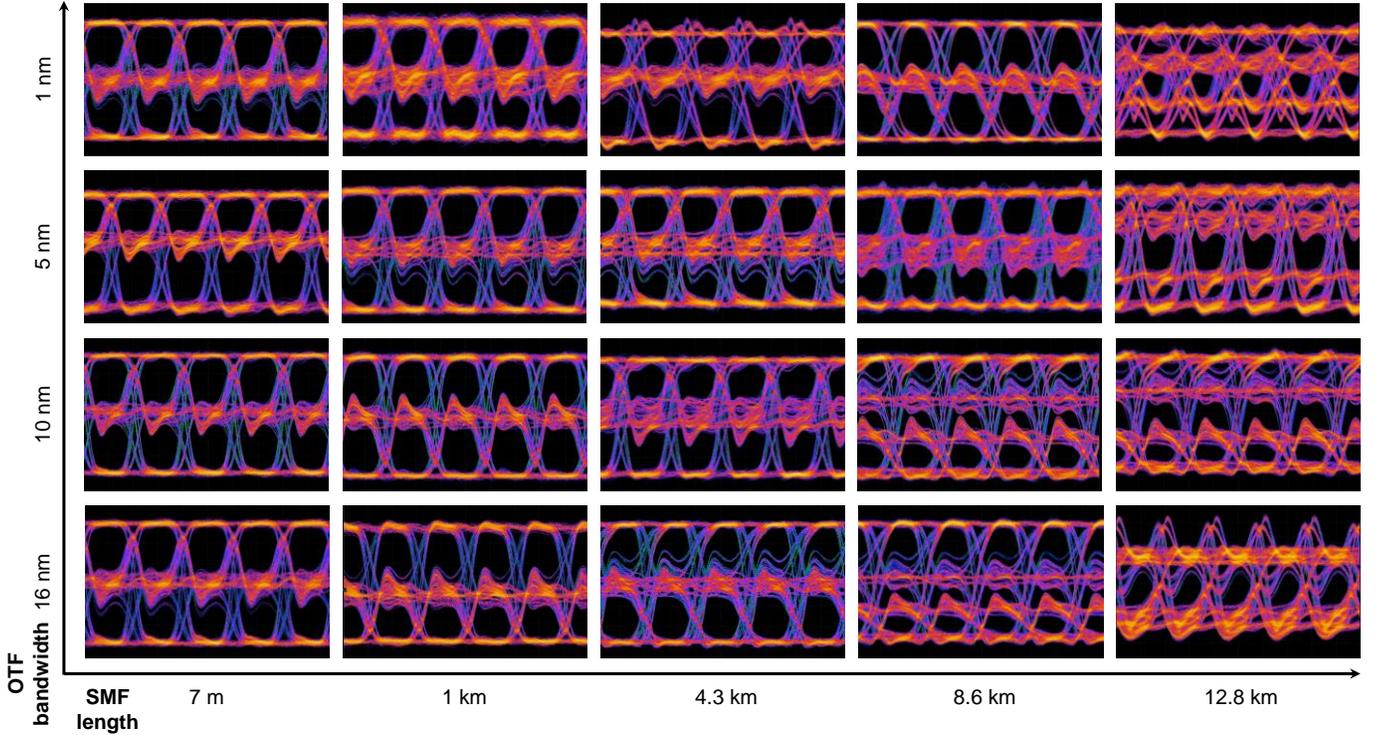

**Fig. 8.** Eye diagrams for polarization encoding at 100 MHz, showing the effect of depolarization for different fiber lengths.

Figure 7 shows the exemplary evolution of the polarization state for four slices from 1570 to 1585 nm, spaced 5 nm apart. As expected, the overall polarization drift increases with the increasing fiber length from Fig. 7(a) to Fig. 7(e). Additionally, the separation of the detected states of polarization on the Poincare sphere increases with increased fiber lengths.

These results have the following implications for using the proposed Ge-on-Si PIN junction, or any other broadband source, for QKD protocols based on polarization encoding. The operation of such a QKD protocol requires alignment of the transmitter polarization with the receiver polarization. Since the polarization evolution as well as the received state of polarization depends on the wavelength and fiber length, this alignment cannot be conducted as for a coherent light source, leading to a QBER penalty. This necessitates to narrow down the emission spectrum of an incoherent light source, at the expense of its emission power delivered to the modulator. Figures 7(f) and 7(g) show the depolarization effect for polarization states being encoded in the **D/A** and **R/L** bases for transmission over a patchcord with a length of 7 m and a fiber spool of 12.8 km, respectively. After polarization alignment at the receiver, we notice a clear recovery of the data signals in both planes for the short distance, while for the long fiber span we cannot recover the transmitted states in both bases anymore and the RF signal power starts to fade at the same time.

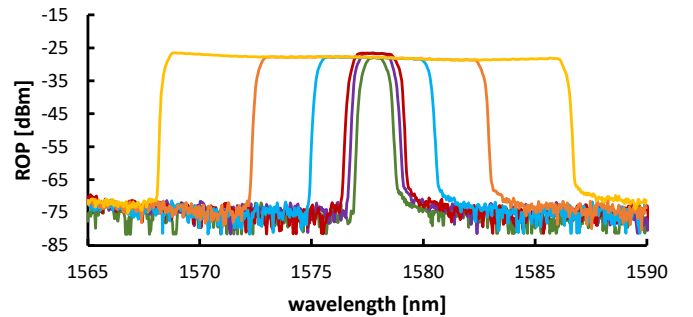

**Fig. 9.** Spectra of the transmitted signal at the classical power level for different OTF bandwidth settings of 1 nm, 1.5 nm, 2 nm, 5 nm, 10 nm and 16 nm.

## VI. Incoherently Seeded QKD Transmitter Employing an I/Q Modulator for State Preparation

We finally evaluated the state preparation through a dual-polarization I/Q modulator with an incoherent light source. Figure 3(b) presents the modified QKD transmitter at Alice, while Bob's receiver was slightly modified to acquire the quantum signal in the **D/A** and **R/L** bases. The EDFA-based incoherent light source is spectrally defined by the OTF and levelled to a mean photon number of $\mu_Q = 0.1$ photons/symbol at the I/Q modulator output through a VOA before the modulator.

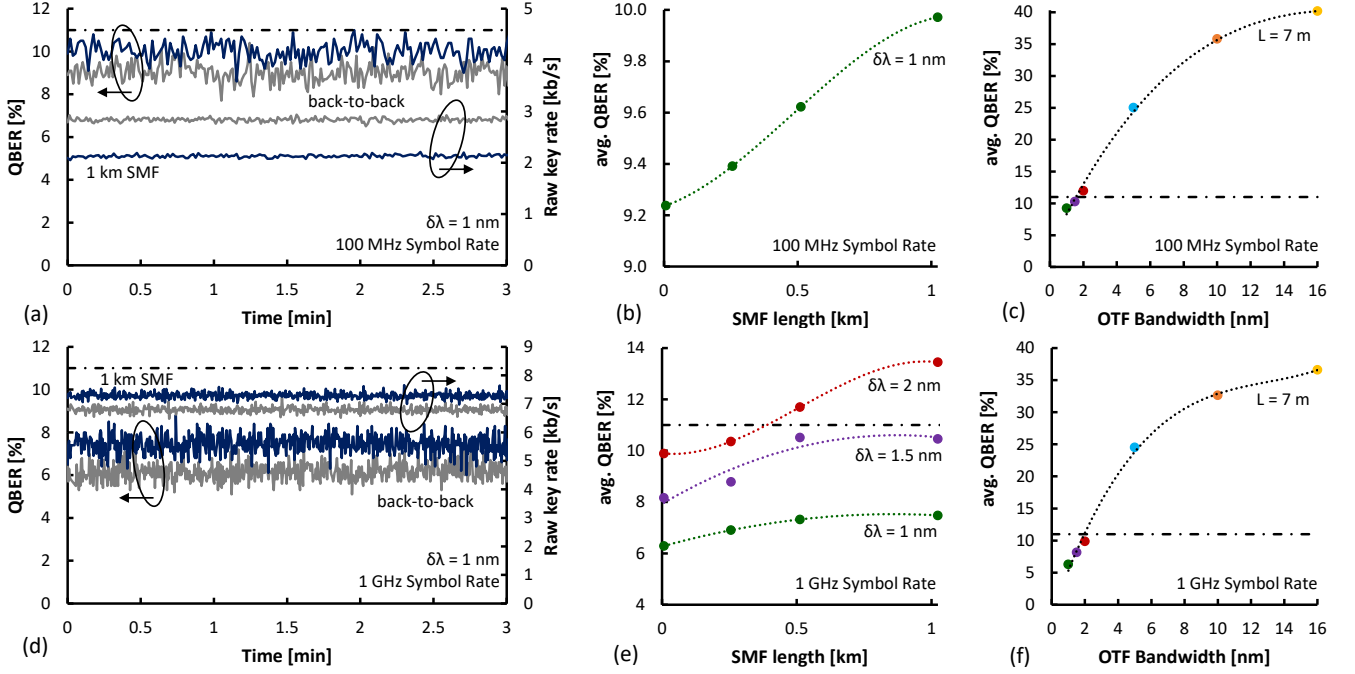

**Fig. 10.** Evolution of the QBER and raw-key rate for δλ = 1 nm and a symbol rate of (a) 100 MHz and (d) 1 GHz for back-to-back (grey) transmission and transmission over 1 km of SMF (blue), with the corresponding dependence of the average QBER on the SMF length shown for symbol rates of (b) 100 MHz and (e) 1 GHz, respectively. Dependence of the QBER in a back-to-back configuration on the optical signal bandwidth δλ for a symbol rate of (c) 100 MHz and (f) 1 GHz.

The employed LiNbO$_3$ dual-polarization I/Q modulator is operated in such way that the child MZMs are biased at their maximum transmission point and serve solely the purpose of balancing the power of both polarization states. We then directly modulate the phase $\varphi$ of one polarization, as described earlier, using an AWG, thereby preparing the transmitted photons in one of the four polarization states **A**, **D**, **R** or **L**. The photons are then transmitted through different optical fibers with lengths up to 1 km before they are detected at Bob's side after manual polarization control. For this second experiment, we resorted to a different pair of InGaAs SPADs, characterized by an efficiency of 10%, a dead time of 25 μs and dark count rates between 500 and 600 counts.

### A. Depolarization of spectrally wide optical signal

To investigate the depolarization effects on data signals, we characterized the signal integrity at a classical transmit power level using balanced receivers. Various settings for the optical bandwidth of the transmitted signals and the SMF length have been investigated, as summarized in Fig. 8 in terms of received eye diagrams for polarization encoding at a symbol rate of 100 MHz. The eye spreads from a clearly delineated three-level signal to four levels once the fiber length increases. Moreover, an increasing optical bandwidth offsets the threshold upon which the depolarization effect becomes visible.

### B. QKD performance

To quantify the depolarization penalty for QKD applications, we have investigated the QBER and raw-key rate accordingly. Measurements have been conducted for symbol rates of 100 MHz and 1 GHz and a temporal filtering of 50% within the symbol period. In the following we only mention the QBER values and raw-key rates on a per-SPAD level, with the values for the second SPAD being nearly identical and differing by less than 5% from the values reported by the other SPAD. This is because the raw-key rates are also reported on a per-basis level due to consecutive measurements conducted with a limited stock of SPAD receivers. This implies that with a fully furnished QKD receiver at Bob, operating four detectors (one dedicated to each of the four transmitted states **R**, **L**, **A** and **D**), one could possibly reach the quadrupled overall key rate.

For the lower symbol rate of 100 MHz and a 1-nm wide slice centered around 1578 nm (Fig. 9) we obtain an average QBER of 9.24% (3σ = 1.45%) for the back-to-back scenario and 9.97% (3σ = 1.36%) for transmission over 1 km with respective raw-key rates of 2.85 kb/s and 2.16 kb/s, respectively (Fig. 10(a)).

For the higher symbol rate of 1 GHz, we are able to reach QBERs of 6.29% ($3\sigma$ = 1.20%) and 7.66% ($3\sigma$ = 1.27%) with corresponding raw-key rates of 6.81 kb/s and 7.38 kb/s for back-to-back transmission and 1 km of fiber, respectively (Fig. 10(d)). These QBERs are well below the threshold of 11%.

The higher key rates and lower QBERs of the 1 GHz system are directly connected to the temporal filtering of 50%, leading to suppressed noise photons and dark counts, thereby advocating the system operating at 1 GHz. On the other hand, since we only post-select our detection events through introducing narrowed detection windows rather than employing pulse carving before signal transmission, we might penalize our key rate due to possible detector saturation effects.

For the remaining investigation concerning various settings of the optical signal bandwidth and transmission reach we will exclusively focus on the QBER since the raw-key rates primarily depend on the optical budget and thereby the loss of the transmission link, rendering it as a less attractive performance indicator.

For the lower symbol rate of 100 MHz, we reach QBERs below 11% for an optical bandwidth of $\delta\lambda$ = 1 nm and all fiber lengths up to 1 km, the latter being the upper limit for our application scenario of intra-datacenter interconnects (Fig. 10(b)). For larger optical bandwidths we were not able to reach QBERs below 11% for a fiber reach exceeding a few meters.

This dependence on $\delta\lambda$ is investigated in Fig. 10(c) for a back-to-back case, which relates to a patchcord with a length of 7 m between the QKD transmitter and receiver. Here, the QBER rises above the threshold of 11% for a bandwidth of 2 nm, reaching an average of 25.03 % ($3\sigma$ = 1.82 %) at 5 nm and 40.21 % ($3\sigma$ = 2.56 %) at 16 nm. We attribute this rapid increase in QBER to an increase in noise photons due to depolarization, causing the signal-to-noise ratio to degrade. Since the balanced receivers employed for the classical characterization do not measure a general increase in background photons due to their architecture, this effect cannot be seen in the eye diagrams in Fig. 8, apart from the fading signal power in Fig 7(g).

Due to the narrower post-selection filtering when operating the QKD system at 1 GHz, the results improve. Here, we obtain QBER values below the threshold of 11% for bandwidths of 1 and 1.5 nm and a transmission reach of up to 1 km, while a bandwidth of 2 nm is compatible with a reach of up to 250 m (Fig. 10(e)). Figure 10(f) summarizes the QBER penalty as function of $\delta\lambda$ for a back-to-back configuration with 7 m of SMF.

## VII. Conclusion

We showed successful raw-key generation at 2.15 kbit/s with a QBER of 7.71 % for a BB84-based QKD transmitter sourced by an incoherent Ge-on-Si light source. Furthermore, we investigated the implications that the broadband emission of this light source has on the transmission over fiber-based optical distribution networks. We found that narrow spectral filtering is necessary to restrict the bandwidth of the emission spectrum to 2 nm, in order to operate an incoherently sourced QKD transmitter below the QBER threshold necessary to generate a secure key. We have further proven the polarization state generation in two bases, as required for the implementation of a polarization-encoded BB84 protocol, by simple phase modulation within a dual-polarization I/Q modulator. This allows for further simplification, eventually enabling, together with the Ge-on-Si light source, an all-silicon QKD transmitter with small form-factor for low-cost applications in the datacenter realm. We expect further improvement on the proposed concept through additional pulse carving before state preparation and monolithic integration on a single photonic chip without the need for fiber coupling between discrete components, which was necessary for our demonstration. Still, some obstacles remain on the path towards an all-silicon QKD transmitter: The polarization multiplexed I/Q architecture itself can be simplified to reduce loss, for example by replacing the nested I/Q with single MZM structures and building on silicon 2D grating couplers for polarization multiplexing. At the same time, the Ge-on-Si PIN junction has to be submitted for further re-design towards a higher light output and narrower spectral characteristics. Altogether, these require engineering efforts on the device level, which we believe can be addressed in the near future.